\documentclass[prl,twocolumn,nofootinbib]{revtex4-1}
\usepackage{graphicx}
\usepackage{color}
\usepackage{dcolumn}
\usepackage{bm}
\usepackage{amssymb}
\usepackage[bottom]{footmisc}
\usepackage{footnote}
\usepackage{lipsum}
\usepackage{gensymb}
\usepackage{wrapfig}
\usepackage{sidecap}
\renewcommand{\thefootnote}{\fnsymbol{footnote}}

\newcommand\blfootnote[1]{%
\begingroup
\renewcommand\thefootnote{}\footnote{#1}%
\addtocounter{footnote}{-1}%
\endgroup}
\begin{document}





\title{Realization of graphene logics in an exciton-enhanced insulating phase}



\author{Kaining Yang,$^{1,2*}$ Xiang Gao,$^{1,2*}$ Yaning Wang,$^{3,4*}$ Tongyao Zhang,$^{1,2*}$ Pingfan Gu,$^{5,6}$ Zhaoping Luo,$^{3}$ Runjie Zheng,$^{7}$ Shimin Cao,$^{7,8}$  Hanwen Wang,$^{3}$ Xingdan Sun,$^{3}$ Kenji Watanabe,$^{9}$ Takashi Taniguchi,$^{10}$ Xiuyan Li,$^{3}$ Jing Zhang,$^{1,2}$ Xi Dai,$^{11,12\dagger}$ Jianhao Chen,$^{7,8,13,14\dagger}$ Yu Ye,$^{5,6\dagger}$ Zheng Vitto Han,$^{1,2\dagger}$}

\affiliation{$^{1}$State Key Laboratory of Quantum Optics and Quantum Optics Devices, Institute of Opto-Electronics, Shanxi University, Taiyuan 030006, P. R. China}
\affiliation{$^{2}$Collaborative Innovation Center of Extreme Optics, Shanxi University, Taiyuan 030006, P.R.China}
\affiliation{$^{3}$Shenyang National Laboratory for Materials Science, Institute of Metal Research, Chinese Academy of Sciences, Shenyang 110016, China}
\affiliation{$^{4}$School of Material Science and Engineering, University of Science and Technology of China, Anhui 230026, China}
\affiliation{$^{5}$Collaborative Innovation Center of Quantum Matter, Beijing 100871, China}
\affiliation{$^{6}$State Key Lab for Mesoscopic Physics and Frontiers Science Center for Nano-Optoelectronics, School of Physics, Peking University, Beijing 100871, China}
\affiliation{$^{7}$International Center for Quantum Materials, School of Physics, Peking University, Beijing 100871, China}
\affiliation{$^{8}$Beijing Academy of Quantum Information Sciences, Beijing 100193, China}
\affiliation{$^{9}$Research Center for Functional Materials, National Institute for Materials Science, 1-1 Namiki, Tsukuba 305-0044, Japan}
\affiliation{$^{10}$International Center for Materials Nanoarchitectonics, National Institute for Materials Science,  1-1 Namiki, Tsukuba 305-0044, Japan}
\affiliation{$^{11}$Materials Department, University of California, Santa Barbara, CA 93106-5050, USA}
\affiliation{$^{12}$Department of Physics, The Hongkong University of Science and Technology, Clear Water Bay, Kowloon 999077, Hong Kong, China}
\affiliation{$^{13}$Key Laboratory for the Physics and Chemistry of Nanodevices, Peking University, Beijing 100871, China}
\affiliation{$^{14}$Interdisciplinary Institute of Light-Element Quantum Materials and Research Center for Light-Element Advanced Materials, Peking University, Beijing 100871, China}


\maketitle
\blfootnote{\textup{*} These authors contribute equally.}

\blfootnote{$^\dagger$Corresponding to: daix@ust.hk, chenjianhao@pku.edu.cn, ye$\_$yu@pku.edu.cn, and vitto.han@gmail.com}

\textbf{For two decades, two-dimensional carbon species, including graphene, have been the core of research in pursuing next-generation logic applications beyond the silicon technology. Yet the opening of a  gap in a controllable range of doping, whilst keeping high conductance outside of this gapped state, has remained a grand challenge in them thus far. Here we show that, by bringing Bernal-stacked bilayer graphene in contact with an anti-ferromagnetic insulator CrOCl, a strong insulating behavior is observed in a wide range of positive total electron doping $n_\mathrm{tot}$ and effective displacement field $D_\mathrm{eff}$ at low temperatures. Transport measurements further prove that such an insulating phase can be well described by the picture of an inter-layer excitonic state in bilayer graphene owing to electron-hole interactions. The consequential over 1 $\mathrm{G\Omega}$ excitonic insulator can be readily killed by tuning $D_\mathrm{eff}$ and/or $n_\mathrm{tot}$, and the system recovers to a high mobility graphene with a sheet resistance of less than 100 $\mathrm{\Omega}$. It thus yields transistors with ``ON-OFF'' ratios reaching 10$^{7}$, and a CMOS-like graphene logic inverter is demonstrated. Our findings of the robust insulating phase in bilayer graphene may be a leap forward to fertilize the future carbon computing.}




 \bigskip

Two-dimensional (2D) materials host great potential in the pursuit of logic applications with emerging quantum degrees of freedom yet smaller sizes other than silicon transistors \cite{Geim_rise_of_Gr, Geim_vdW_2013, MoS2_CPU_NC_2017}. Among the 2D materials, graphene is known to exhibit fascinating electronic properties, especially the high mobility that is crucial to reduce power consumption at the ``ON'' state. However, the fact that graphene lacks an insulating ``OFF'' state, due to its gapless semi-metallic nature, has hindered its applications. Efforts have been devoted to open a band gap in graphene for over two decades, via both physical and chemical manners \cite{G_NanoMesh_NN_2010, Philip_PRL_2007}. Unfortunately, they either ended up with a severely degraded electrical conductivity at the ``ON'' state such as the meshed graphene in a nano scale \cite{G_NanoMesh_NN_2010}, or turned out to be gapped in an impractical narrow doping range, such as those found in the case of a bilayer graphene at the charge neutrality subjected to vertical electrical fields \cite{WangFeng_BLG_Gap_Nature, LiJing_Jun_Zhu_PRL_2018}.


   \begin{figure*}[ht!]
   \includegraphics[width=0.88\linewidth]{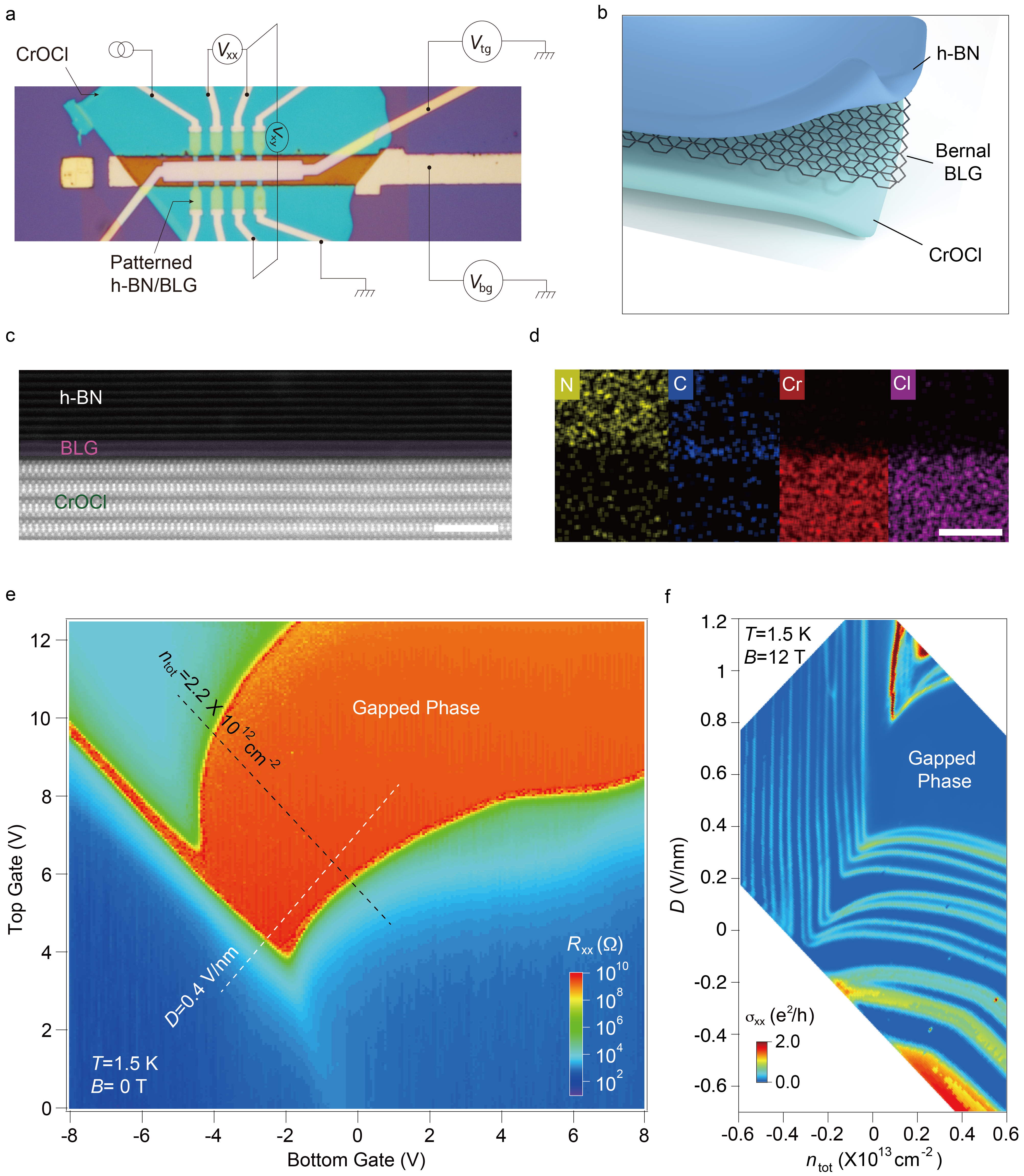}
   \caption{\textbf{Characterization of h-BN/BLG/CrOCl heterstructures.} (a) Optical micrograph image of a typical h-BN/BLG/CrOCl sample, with a cartoon illustration of the heterostructure shown in (b). (c) HAADF-STEM image of the cross-section of a typical sample shown in (a), with the EDS mapping of N, C, Cr and Cl elements given in (d). Scale bars in (c) and (d) are 2 nm and 5 nm, respectively. (e) Color map of a dual gate scan of field effect in a typical sample, measured using DC Ohm meter at $T$= 1.5 K and $B$=0 T. (f) $\sigma\mathrm{_{xx}}$ scanned in the same gate range as in (e), and re-plotted in the parameter space of $D$ and $n\mathrm{_{tot}}$, data measured using AC lock-in at $T$= 1.5 K and $B$= 12 T.
   }
   \end{figure*}

Except for the band insulator or disorder-induced gaps mentioned above, it was proposed that, rather than single particle electrons, a bosonic insulating ground state with electron-hole pairing interactions can occur spontaneously with a binding energy $\varepsilon\mathrm{_{b}}$ overcoming the gap $\varepsilon\mathrm{_{g}}$ between valence and conduction bands, expected preferably in semimetals with zero gaps \cite{Mott_1961, Knox_1963}. Such an insulator is regarded as an excitonic condensation, which is considered to be a superfluid in analogue to a Bardeen-Cooper-Schrieffer (BCS) superconductor \cite{Keldysh_1964, Jerome_1967, Halperin_RMP_1968, Snoke_Science_2002}. This excitonic condensation was first reported in a number of vertical double layer quantum wells \cite{Blatt_PR_1962, Lozovik_1975, Zhu_PRL_1995, Eisenstein_Nature_2004, High_Nature_2012, DuRuiRui_NC_2017}, and recently revisited in 2D material double layer systems \cite{Leo_NatPhys_2017, Xiaomeng_NP_2017, MacDonald_PRL_2018, Mak_Nature_2019, ShanJie_arXiv_2021_Capacitance}. Other than the indirectly-coupled double layers spaced by a finite distance, it is intriguing to realize excitonic insulator directly in low dimensional electronic systems by either reducing the $\varepsilon\mathrm{_{g}}$ or by enhancing the Coulomb interaction. Indeed, a series of recent theories and experiments suggest possible gapped excitonic phases in such as Sb nano-flakes \cite{WangXiaoLin_NanoLett_2019}, single crystals of Ta$_{2}$NiCh$_{5}$ (Ch=S, Se and Te) \cite{Ta2NiSe5_NC_2017} and 1T-TiSe$_{2}$ \cite{Kogar_Science}, as well as transition metal dichalcogenides monolayer WTe$_{2}$ \cite{WuSanFeng_arXiv_2020} and T' phase MoS$_{2}$ \cite{Topological_TPhase_MoS2_2020_NN}. Meanwhile, introducing or enhancing such electron-hole interaction induced gap in graphene in a controlled manner, and in a wide doping range, will be of potential for nanoelectronic applications, which however remains largely unexplored thus far.




In this work, we study the case of Bernal-stacked bilayer graphene (BLG). It is predicted that in the existence of strong Coulomb interaction at the charge neutrality in a BLG, the excitonic instabilities will be triggered via the bound of electron-hole pairs, and the system undergoes a phase transition to an excitonic insulator at the ground state \cite{BLG_Exciton_Insulator_PRB_2012, BLG_Exciton_Insulator_PRB_2006}. To realize the regime of strong Coulomb interaction, we design the system by bringing BLG in touch with CrOCl, an antiferromagnetic insulator that serves as a reservoir of electrons due to interfacial coupling, with a built-in electrical field triggered at certain total doping and displacement field induced by the dual gates \cite{Submitted}. Consequently, a strong gapped behavior is observed in the h-BN/BLG/CrOCl heterostructure with sheet resistance reaching G$\Omega$ in a wide range in the parameter space of total electron doping $n_\mathrm{tot}$, effective displacement field $D_\mathrm{eff}$, magnetic field $B$, as well as temperature $T$. We attribute this insulating phase to an Coulomb interaction enhanced inter-layer excitonic state, which yields transistors with ``ON-OFF'' ratios reaching 10$^{7}$, and a CMOS-like graphene logic inverter is further demonstrated. Our results pave the way for the engineering of such an insulating state, which may be expanded to a broader library of materials and may find applications in future quantum electronics.
 



\textbf{Characterizations of h-BN/BLG/CrOCl heterostructures.} 

Bernal-stacked BLG, thin CrOCl flakes, and encapsulating hexagonal boron nitride (h-BN) flakes were exfoliated from high-quality bulk crystals and stacked in ambient condition using the dry transfer method.\cite{Lei_Science} Electrodes with edge-contacts to Hall bars of the vertically assembled van der Waals heterostructures were patterned using standard electron beam lithography. A dual-gate configuration is used to capacitively induce carriers $n\mathrm{_{tot}}$ and to define a vertical displacement field $D$ from both top and bottom gates. The optical micrograph of a typical sample is shown in Fig. 1a, with the cartoon view of the h-BN/BLG/CrOCl illustrated in Fig. 1b. Atomic resolution of the cross section of a typical heterostructure can be seen in the high-angle annular dark-field scanning transmission electron microscopy (HADDF-STEM) image in Fig. 1c, with the corresponding electron dispersive spectroscopy (EDS) mapping shown in Fig. 1d. We cooled down the samples to a base temperature of 1.5 K, and measured the longitudinal channel resistance $R\mathrm{_{xx}}$ as a function of top gate $V\mathrm{_{tg}}$ and bottom gate $V\mathrm{_{bg}}$, as shown in Fig. 1e. It is seen that $R\mathrm{_{xx}}$ can be tuned via gates from less than 10$^{2}$ $\Omega$ to a highly insulating state reaching the order of 10$^{10}$ $\Omega$. Such a strong gapped state is attributed to an excitonic insulator, as will be discussed in the coming text, which exists in a very wide range of gate voltages. 

With the dual gate configuration, capacitively induced carrier density through the two gates can be decoupled from the displacement field, with the total carrier density defined as $n\mathrm{_{tot}}=(C_\mathrm{tg}V_\mathrm{bg}+C_\mathrm{bg}V_\mathrm{bg})/e-n_{0}$, and the applied average electric displacement field $D=(C_\mathrm{tg}V_\mathrm{tg}-C_\mathrm{bg}V_\mathrm{bg})/2\epsilon_{0} - D_{0}$, where $C_\mathrm{tg}$ and $C_\mathrm{bg}$ are the top and bottom gate capacitances per area, respectively. $n_{0}$ and $D_{0}$ are residual doping and residual displacement field, respectively. In order to extract the $n_{0}$ and $D_{0}$, we measured a typical h-BN/BLG/CrOCl heterostructure sample at $B$= 12 T and $T$=1.5 K, where Landau Levels (LLs) can be developed outside the gapped state, with the transverse resistance quantized at each filling fractions. One can thus define the $n\mathrm{_{tot}}$=0 using the LLs, and hence $n_{0}$ can be obtained, same as previously reported \cite{Submitted}. Since $D$ is a relative value, we chose $D_{0}$ at the $n\mathrm{_{tot}}$=0 line with $V_\mathrm{bg}=0$, for simplicity. 

As shown in Fig. 1f, longitudinal conductivity $\sigma_\mathrm{xx}=\frac{R_\mathrm{xx}}{(R_{\mathrm{xx}}^{2}+R_{\mathrm{xy}}^{2})}$ at $B$ = 12 T and $T$=1.5 K is plotted in the parameter space of $D$ and $n\mathrm{_{tot}}$. Hole and electron sides show distinctive features of Landau quantizations, with a crossover from straight stripes to cascades-like stripes as the total doping switches from hole to electron sides, very similar to the results obtained from the MLG/CrOCl samples except for the large gapped area in the BLG ones \cite{Submitted}. Here, we adopt the definition from the MLG/CrOCl case that the doping area, with distorted LLs more tunable with $D$ instead of $n\mathrm{_{tot}}$, is named as a strong-interfacial-coupling (SIC) phase \cite{Submitted}. More discussions on the magnetic field dependence (measured along the black dashed line in Fig. 1e) of LLs in the BLG/CrOCl samples can be found in Extended Data Figures 1-2. Full degeneracy lifting with quantized plateaux obtained in $R_\mathrm{xy}$ was seen at each integer filling fractions from $\nu$=-2 to -10. This speaks the high mobility of the graphene in the conventional phase in the BLG/CrOCl heterostructure. Indeed, by fitting of the hole-side part of field-effect curve at zero magnetic field (Extended Data Fig. 3), hole carrier mobility is estimated to be a few 10$^{3}$ cm$^{2}$V$^{-1}$s$^{-1}$. In the following main text, we will mainly focus on the gapped region.

   \begin{figure*}[ht!]
   \includegraphics[width=0.88\linewidth]{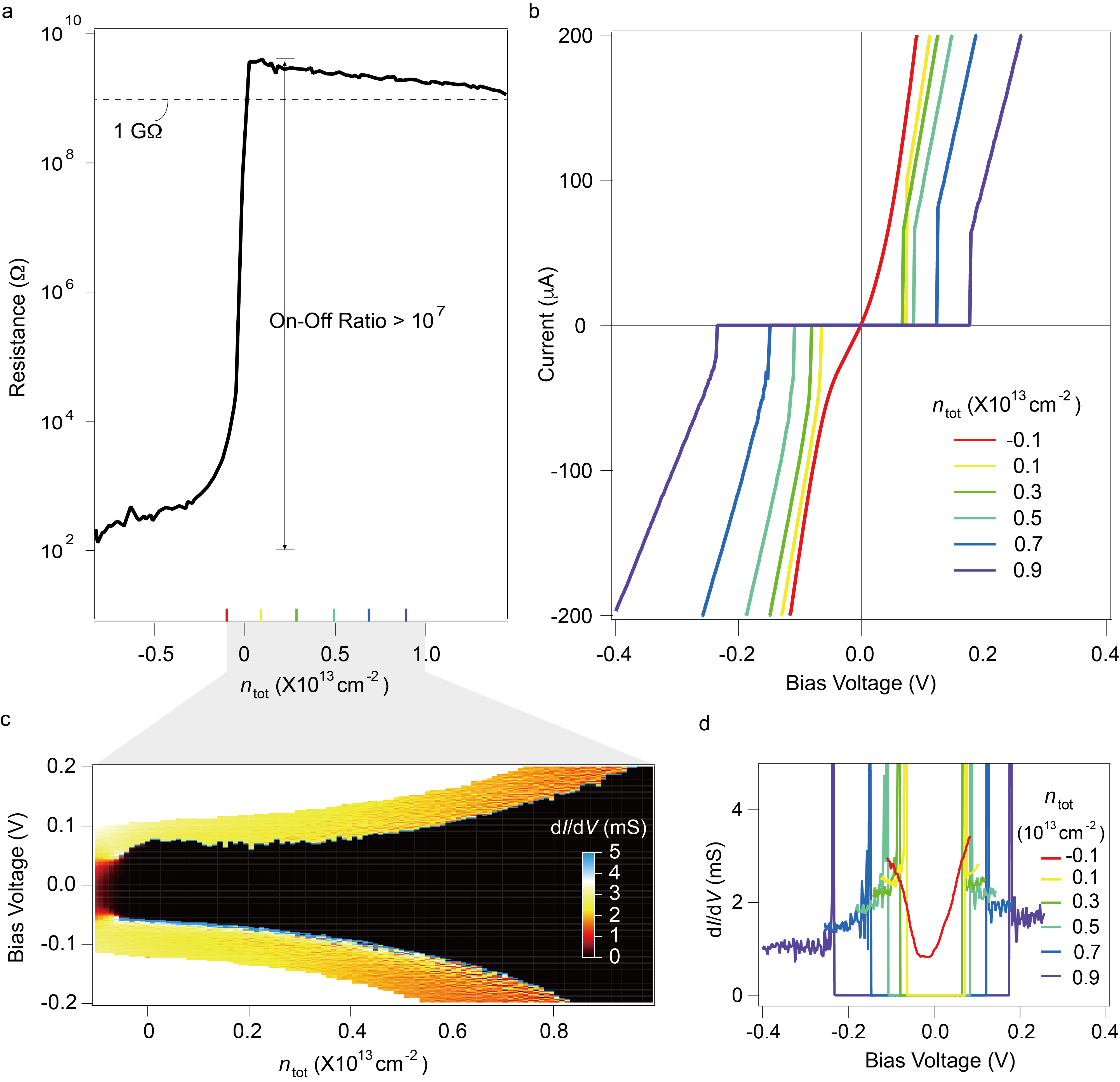}
   \caption{\textbf{Gate tunable semi-metal to insulator transition.} (a) Line profile of $R\mathrm{_{xx}}$ at $D=0.4$ V/nm, in the range of  $-0.8 \times  10^{13}$ cm$^{-2}$ $< n\mathrm{_{tot}} < 1.5 \times  10^{13}$ cm$^{-2}$. (b) $I$-$V$ curves at various $n\mathrm{_{tot}}$, marked as color ticks on the axis in (a). (c) Mapping of $I$-$V$ curves along $D=0.4$ V/nm, in the range of  $-0.1 \times  10^{13}$ cm$^{-2}$ $< n\mathrm{_{tot}} < 1.0 \times  10^{13}$ cm$^{-2}$. (d) Differential conductance d$I/$d$V$ as a function of bias voltage obtained from the $I$-$V$ curves shown in (b).}

   \label{fig:fig2}
   \end{figure*}

It is noticed that, at zero magnetic field, a resistive peak is also found at the charge neutrality in MLG/CrOCl samples, whose absolute value of resistance is a few tens of k$\Omega$, and located in a narrow range near the charge neutrality \cite{Submitted}. However, in the BLG/CrOCl scenario, the resistive peak reaches a value of a few G$\Omega$, and expands to an area across difference in displacement field of $\delta D> $ 0.5 V/nm, starting from $n\mathrm{_{tot}} >$0, as can be seen in Fig. 1e-f. We performed dual-gated mappings of channel resistance at various temperatures, as shown in Extended Data Figure 4. It is seen that the gapped region shrinks its doping area with increasing temperature, with the resistance value decreasing as well. The resistive peak prevails at 100 K at the charge neutrality, with an amplitude of about a few M$\Omega$, still much larger a value as compared to many previously reported gapped states in graphene. 


\begin{figure*}[ht!]
 \includegraphics[width=0.90\linewidth]{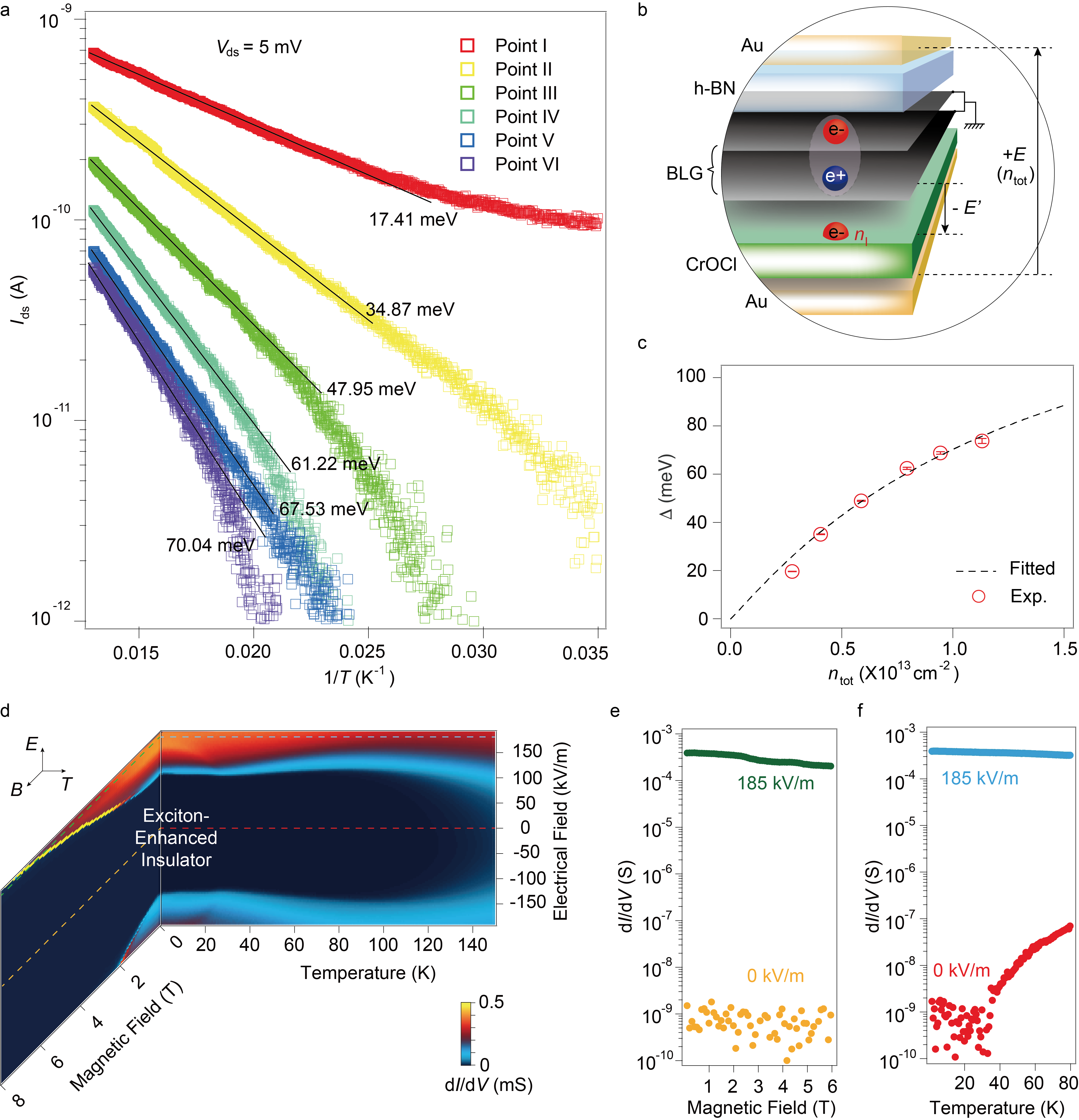}
 \caption{\textbf{Characteristics of the EIS in BLG/CrOCl heterostructures.} (a) $I_\mathrm{ds}$ as a function of 1$/T$ in a semi-log  plot. Thermal activation gaps extracted from each curve are labelled. (b) Art view of the electron-hole paring thanks to the built-in electrical field at the graphene/CrOCl interface, while trapped electrons $n_\mathrm{I}$ in CrOCl does not contribute to transport and the $n\mathrm{_{tot}}$ is defined by the two gates. (c) Thermal activation gaps extracted from curves in (a), as a function of $n\mathrm{_{tot}}$. Error bars are  standard deviations of the linear fit in (a). The dashed line is a theoretical fit using Eq. (1). (d) Differential conductance d$I/$d$V$ as a function of bias voltage (renormalized to in-plane electrical field) in the parameter space of $B$ and $T$. Data were measured at point-II, as defined in Extended Data Figure 6. (e) and (f) are line cuts along dashed lines in (d).}
 
 \label{fig:fig3}
 \end{figure*}

\textbf{Gate tuned semi-metal to extremely-insulating-state transition in BLG.} 
\\
To analyse the observed extremely-insulating-state (EIS), we first extract the line profiles of $R_\mathrm{xx}$ at a fixed $D$ ($D$=0.4 V/nm), indicated by the white dashed line in Fig. 1e. A contact resistance of about 1600 $\Omega$ is subtracted. As can be seen from Fig. 2a, $R_\mathrm{xx}$ of the BLG/CrOCl heterostructure is tuned from 10$^{2}$ $\Omega$ at the hole side (i.e., the conventional phase of BLG) to over 1 G$\Omega$ at the electron side (i.e., the SIC-phase, notice that the electron side denotes $n\mathrm{_{tot}} >0$), with an ``ON/OFF'' ratio reaching 7 orders of magnitudes over the doping range of $-0.8 \times  10^{13}$ cm$^{-2}$ $< n\mathrm{_{tot}} < 1.5 \times  10^{13}$ cm$^{-2}$, which is, to our knowledge, the record high value reported.

One of the typical characteristics of the observed EIS is the ``critical voltage'' $V\mathrm{_{C}}$, at which the EIS breaks down and turns into a ``normal'' metallic state. To verify it, we performed four-probe $I$-$V$ curve measurements at several values of $n\mathrm{_{tot}}$, as shown in Fig. 2b. Indeed, it is seen that the system exhibits ``gapped'' region as negligible source-drain current can be obtained within $\pm V\mathrm{_{C}}$, which is a perfect insulator behavior. Above $V\mathrm{_{C}}$, the system turns into a metallic state, with resistances of less than 1 k$\Omega$. This could be an evidence of an excitonic insulator, in which electron-hole pairs break into single-particle charge carriers at a threshold bias voltage $V\mathrm{_{bias}}$. Characteristic hysteresis (the trapping-retrapping-like behavior in the hysteresis of $I$-$V$ curves are only seen in such as either BCS superconductors, or in ``super-insulating'' reported in a granular superconducting film \cite{Baturina_Nature}) can be seen in the $I$-$V$ curves (Supplementary Figure 5b), which rules out the existence of band insulator in our system. Similar $I$-$V$ curves were reported in ultra-clean suspended BLG with a spontaneous gap opening at the narrow doping range around charge neutrality \cite{Baowenzhong_PNAS}.

Fig. 2c shows a color map of differential conductance d$I/$d$V$-$V\mathrm{_{bias}}$ curves along $D=0.4$ V/nm, in the range of $-0.1\times 10^{13}$ cm$^{-2}$ $< n\mathrm{_{tot}} < 1.0\times 10^{13}$ cm$^{-2}$, with lines cuts of d$I/$d$V$-$V\mathrm{_{bias}}$ shown in Fig. 2d. In general, $V\mathrm{_{C}}$ is monotonously increasing with increasing $n\mathrm{_{tot}}$. This behavior is an indication that the total density of electron-hole excitons is increasing upon increasing $n\mathrm{_{tot}}$, giving rise to an enhanced binding energy, which corresponding to an elevated $V\mathrm{_{C}}$. Indeed, increasing $n\mathrm{_{tot}}$ means an enhanced built-in electrical field at the graphene/CrOCl interface (Fig. 3b), and thus more available states are induced in the CrOCl, leading to an increase of inter-layer electron-hole interaction, hence enhanced $\varepsilon\mathrm{_{B}}$ \cite{BLG_Exciton_Insulator_PRB_2006}. It is noticed that the $V\mathrm{_{C}}$ in $I$-$V$ curve in the studied system is almost linearly dependent on the distance of sample electrodes, as shown in Extended Data Figure 5. It thus suggests that the critical bias voltage in our present system actually corresponding to the break down voltage of electron-hole excitons over a certain distance ($i.e.$, an in-plane electrical field), rather than an energy gap as in the $I$-$V$ spectroscopy of, for example, a BCS superconductor. Nevertheless, this $V\mathrm{_{C}}$, or, more precisely, the critical in-plane electrical field $E\mathrm{_{C}}$ should be proportional to the binding energy $\varepsilon\mathrm{_{B}}$, as both of the two measures are indications of the coupling strength of the electron-hole excitons.



To further quantitatively clarify the observed EIS in BLG/CrOCl, we carried out temperature scans of the sample source-drain current $I\mathrm{_{ds}}$ at different $n\mathrm{_{tot}}$ (Fig. 3a) along with the charge neutrality defined in the dual-gate map at $B=0$ and $T=80$ K, marked as points I-VI, shown in Extended Data Figure 6. By choosing the 6 points at the charge neutrality, we assume that as shown in Fig. 3b, there are interfacial states $n\mathrm{_{I}}$ appearing in the CrOCl. As is discussed in our electrostatic model (Methods), $n\mathrm{_{I}}$ could be impurity states, or, alternatively, some peculiar buried surface state that is very close to the surface of CrOCl (further experiments such as optical spectroscopy will be needed to justify its origin). In this peculiar case, owing to the built-in electric field at the graphene/CrOCl interface, the hybridization gap at the Fermi level of BLG will be largely enhanced. Coulomb screening can then be significantly reduced and electron-hole interaction can be strong enough to form exciton pairing at low temperatures.

\begin{figure*}[ht!]
 \includegraphics[width=0.90\linewidth]{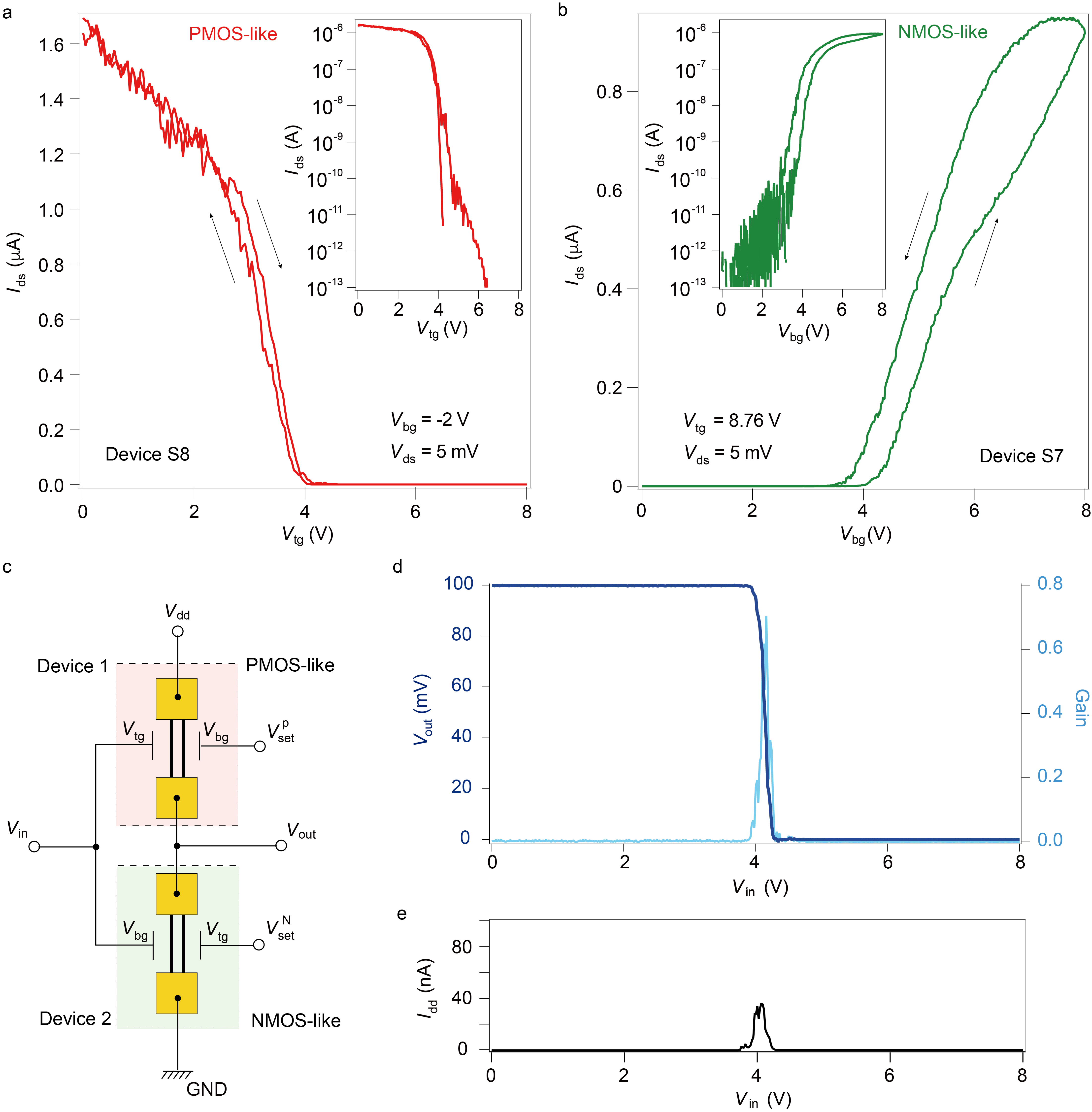}
 \caption{\textbf{Demonstration of the graphene CMOS inverter.} (a)-(b) PMOS- and NMOS-like field effect curves in the same gate range, swept along dash lines in Extended Data Figure 8a-b. Inset of each shows the log scale of the same data. (c) Schematic picture of the BLG/CrOCl CMOS logic inverter. (d) The performance of a typical graphene inverter. (e) Current flowing in the graphene inverter as a function of input voltage during working.}
 
 \label{fig:fig4}
 \end{figure*}

The $I\mathrm{_{ds}}$-$T^{-1}$ curves for the above mentioned 6 points are obtained using DC 2-probe measurement with a fixed $V\mathrm{_{bias}}$=5 mV, as plotted in Fig. 3a. The thermal activation gaps (defined as $I\mathrm{_{ds}}\propto \mathrm{Exp}(-\frac{\Delta}{2KT})$) of each curve are then extracted in Fig. 3c, with each corresponding $n\mathrm{_{tot}}$ calculated from their gate voltages (Extended Data Figure 6). The gap size $\Delta$ in our system can be estimated as a function of $n\mathrm{_{tot}}$ \cite{BLG_Exciton_Insulator_PRB_2006},

\begin{equation}
\Delta\approx \frac{e^{2}L^{2}n\mathrm{_{tot}}}{2C_{b}}\left [ 1+\frac{\Lambda\hbar^{2}\nu^{2}\pi|n\mathrm{_{tot}}|}{\gamma_{1}^{2}} - \Lambda \ln\left (\frac{\hbar\nu\sqrt{\pi|n\mathrm{_{tot}}|} }{\gamma_{1}} \right )\right ]^{-1}
\end{equation}

\noindent where $C_{b}=\epsilon_{r}\epsilon_{0}L^{2}/c_{0}$ (with $\epsilon_{r}$ the BLG dielectric constant, $\epsilon_{0}$ the permittivity of free space, and $c_{0}$ the interlayer distance) is the capacitance of a BLG of area $L^{2}$, $\nu$ is the Fermi velocity, and $\gamma_{1}$ is the interlayer coupling strength. $\Lambda=e^{2}L^{2}\gamma_{1}/(2\pi\hbar^{2}\nu^{2}C_{b})$ is a dimensionless parameter describing the Coulomb screening in the BLG. A best fit of the experimental data in Fig. 3c using Eq. (1) is shown by the black dashed line, yielding fitting parameters of $\nu$=10$^{6}$ m/s, $\gamma_{1}$=0.168, and $\epsilon_{r}$=2.8, whose values are all within reasonable ranges.


For a typical $V\mathrm{_{bg}}$ and $V\mathrm{_{tg}}$ which correspond to point-II in Fig. 3a, we performed the d$I/$d$V$ versus in-plane electrical field $E$ scans in the parameter space of $B$ and $T$, as shown in the phase diagram in Fig. 3d. It is seen that the zero-biased insulating phase can be killed at high temperatures, while the critical in-plane electrical field $E\mathrm{_{C}}$ gets enhanced with increasing $B$. Up to a maximum $B$= 8 T, the $E\mathrm{_{C}}$ is monotonously enhanced, in agreement with a scenario of spinless excitons \cite{SpinlessEI_PR_1968, Elliott_1960}. Line profiles along dashed lines in (d) are shown in (e) and (f), which illustrate d$I/$d$V$ versus temperature or magnetic field below or above the $E\mathrm{_{C}}$. Clearly, when at above $E\mathrm{_{C}}$, the BLG restores back to a semi-metallic state with conductance at the order of mS, and quantum oscillations are seen in neither the  insulating nor the normal semi-metallic phases, in agreement with previously reported excitonic insulator \cite{YuanCao_NJP_2018}. Data obtained from different electrodes at different dopings show similar results, as shown in Extended Data Figure 7. 

It is noteworthy that, compared to the ideal exciton insulator system where the particle numbers of both the electrons and holes residing on two different layers are conserved at the single-particle level, in the BLG systems under finite displacement field there are always finite single-particle tunnelling terms which mix the electron and hole states on different layers and open a ``hybridisation gap''. As shown in the Methods section, we developed an electrostatic model, which introduces interfacial states $n\mathrm{_{I}}$ and self-capacitance of these states $C\mathrm{_{0}}$. It turns out that the observed experimental features in the SIC-phase can be understood by the fact that the $D$ is correlated to chemical potential $\mu$, while the  $n\mathrm{_{I}}$ very close to the BLG plays an role of amplified displacement field that largely enhances (estimated to be one order of magnitude higher than state-of-the-art techniques) the opening of the ``hybridisation gap''. Formation of excitons are expected here due to the strong Coulomb interaction between the electrons and holes, which would manifest itself by the dramatic enhancement of the insulating behavior in the low temperature. As shown in the red dotted curve in Fig. 3f (and Extended Data Figure 7), we do see such trends of drastic drop (at a critical temperature of about 40 K in the red dotted curve in Fig. 3f) of zero-biased differential conductance as the temperature is lowered, which can be explained as a crossover of the gap nature from the hybridisation dominant type in the high temperature to exciton dominant type at low temperatures. However, exact values of exciton binding energy $\varepsilon\mathrm{_{B}}$ requires further investigations such as optical measurements \cite{Luoie_NL, Louie_PRL, JuLong_Sci}.


\bigskip
\textbf{CMOS-like graphene inverter based on the metal-to-EIS phase transition.} 
\\Interestingly, by tuning the doping-driven phase transition from metal-to-EIS, one can obtain both PMOS- and NMOS-like behaviors in the BLG/CrOCl systems in a specific gate range. Taking samples Device-7 and Device-8 for example, as shown in Extended Data Figure 8, the ``ON'' and ``OFF'' state can be out-of-phase when scanned along the dashed lines in the two different samples. More specifically, by setting $V\mathrm{_{bg}}$ at -2 V in Device-8 (setting $V\mathrm{_{tg}}$ at +8.76 V in Device-7), and scan $V\mathrm{_{tg}}$ ($V\mathrm{_{bg}}$) in the range of 0 to 8 V, a PMOS-like (NMOS-like) field-effect curve can be realized, as shown in Fig. 4a and 4b, respectively. Log scale plot of each curve is shown in their insets, indicating an ``ON/OFF'' ratio of 6-7 orders of magnitudes. In the measurements of field-effect curves, $V\mathrm{_{ds}}$ was set to be 5 mV. 

Thanks to the EIS gap in a wide range of doing, the NMOS and PMOS types field-effect curves can now be obtained readily in graphene. One then can design a logic inverter out of the two BLG/CrOCl devices, as illustrated in Fig. 4c (see also Extended Data Figure 8e-f). The diagram of the BLG/CrOCl logic inverter is similar to a standard Si CMOS inverter, but two extra setting voltages ($V\mathrm{_{set}^{P}}$ and $V\mathrm{_{set}^{N}}$) are needed to maintain the shape of the desired field-effect curves. $V\mathrm{_{dd}}$ denotes the supply voltage ($i.e.$, $V\mathrm{_{ds}}$ in the previous conventions), and $V\mathrm{_{in}}$ is the input voltage of the inverter, which is sent to $V\mathrm{_{tg}}$ and $V\mathrm{_{bg}}$ for each device, as shown in Fig. 4c. The performance of such a typical graphene CMOS inverter is shown in Fig. 4d, with a $V\mathrm{_{dd}}$=0.1 V in our measurements. The output voltage $V\mathrm{_{out}}$ is identical to $V\mathrm{_{dd}}$ and flipped to zero at a threshold voltage of about 4 V, yielding a gain of about 0.7. During the working process of the graphene CMOS-like inverter, a maximum $I\mathrm{_{dd}}$ of about 40 nA was seen (Fig. 4e), corresponding to a power consumption of 4 nW. An NAND logic gate was further demonstrated as shown in Extended Data Figure 9.








In conclusion, we have designed a hybrid system with Bernal-stacked BLG interfaced with an antiferromagnetic insulator CrOCl. Thanks to the peculiar graphene/CrOCl interfacial coupling, a strong effective built-in electric field can be established. It thus reduces significantly the Coulomb screening at a wide range of $n\mathrm{_{tot}}$ and $D$ around the charge neutrality, forming a robust insulting state. Transport measurements further revealed an extremely insulating state with resistance exceeding 1 G$\Omega$, $I$-$V$ curves with critical voltage (or defined as in-plane electrical fields), and the monotonously increased $B$-field dependence of such insulating phase. All these evidences point to the existence of a Coulomb interaction enhanced excitonic insulating phase in the studied system. The wide doping range of such a gap further allows us to design and realize a graphene CMOS logic inverter with a gain of about 0.7 and power consumption at the order of 4 nW. Our finding of the robust insulating state in graphene thus paves the way for the future engineering of exotic quantum electronic states in a broader range of material base.

\section{Methods}

$\textbf{Sample fabrications and characterizations.}$ The CrOCl/bilayer-graphene/h-BN heterostructures were fabricated in ambient conditions using the dry-transfer method, with the flakes exfoliated from high-quality bulk crystals. CrOCl bulk crystals were grown via a chemical vapour transport method. Thin CrOCl layers were patterned using an ion milling with Ar plasma, and dual-gated samples are fabricated using standard e-beam lithography (Zeiss Sigma300 + Raith ELPHY Quantum). A Bruker Dimension Icon atomic force microscope was used for thicknesses and morphology tests. The electrical performances of the devices were measured using a Oxford TeslaTron with a base temperature of 1.5 K and a superconducting magnet of 12 T maximum. A probe station (Cascade Microtech Inc. EPS150) is used for room temperature electrical tests. For AC measurements, Standford SR830 lock-ins were used at 17.77 Hz to obtain 4-wire resistances, in constant-current configuration with a 100 M$\Omega$ AC bias resistor. For DC measurements, we used Keithley 2636B multimeters for high precision current measurements, and Keithley 2400 source meters for providing gate voltages. The STEM and EDS investigations were conducted using a double aberration corrected FEI Themis G2 60-300 electron microscope equipped with a SuperX-EDS detector and operated at 300kV.

$\textbf{Electrostatic model.}$
To understand the h-BN/BLG/CrOCl system equipped with two gates, we adopt the following key assumptions in our electrostatic model. First, we assume that there exists interfical states $n\mathrm{_{I}}$ near the interface between CrOCl and BLG, with an average distance between these states and BLG defined as $d$.  The density of states (DOS) of the above mentioned interface states is $\rho_I$, which should be much larger than the low energy DOS of BLG, denoted as $\rho_{\mathrm{BLG}}$. Second, as shown in Extended Data Figure 10a, the entire system containing the interface and BLG can be viewed as one capacitance with effective capacity denoted as $C_{\mathrm{eff}}$, , where the top-gate and back-gate dielectric capacitances are comparable. The total charge of the interface and BLG can be determined by the $C_{\mathrm{eff}}$ and the total voltage as $Q_{\mathrm{tot}} = C_{\mathrm{eff}}(V_\mathrm{tg} + V_\mathrm{bg})$, with $V_\mathrm{tg}$ and $V_\mathrm{bg}$ being the voltage at top and bottom gates, respectively. However, how the above total charge distributes between the interface states and BLG states will be determined by the detailed self-balance of charges between the two. Apparently, the relative energy difference between the bottom energy of the interface states and the charge neutrality point (CNP) energy of BLG is simply tuned linearly by the back gate voltage $V_\mathrm{bg}$, as illustrated in Extended Data Figure 10b.

Following the experimental conventions, we define two voltages,
\begin{equation}
  V_1 = (V_\mathrm{tg}+V_\mathrm{bg})/2; \hspace{3em} V_2 = (V_\mathrm{tg}-V_\mathrm{bg})/2
\end{equation}

Here, $V_1$ and $V_2$ are variables that plays the role of $n_\mathrm{tot}$ and $D$ in the main text, but simplified for the following analysis. 

From the above assumptions, we have
\begin{equation}
  n_{\mathrm{tot}} = n_{\mathrm{BLG}} + n_\mathrm{I} = 2 C_{\mathrm{eff}} V_1/e
\end{equation}

Considering the self capacitance of the interface $C_0$, the occupation numbers of the interface states can then be determined as,
\begin{equation}
  n_\mathrm{I} = e\rho_\mathrm{I} \left[ a (V_1 - V_2) + \mu / e - \frac{e n_I}{C_0} \right] 
\end{equation}
\noindent where $\mu$ is the chemical potential of the Fermi surface.

Notice that, in Eq.(4), the term $a(V_1-V_2)$ implies that the phase boundary between conventional phase and the SIC-phase will be located at a $V_1-V_2=0$ line, as indeed is approximately seen in the resistance map at 12 T and 1.5 K plotted in the $V_1$-$V_2$ space, shown in Extended Data Figure 10c (black dashed line).

Solving Eq.(4), one obtains,
\begin{equation}
n_\mathrm{I} = \frac{e \rho_\mathrm{I} a (V_1 - V_2) + \rho_\mathrm{I} \mu}{1 + e^2 \rho_\mathrm{I}
  / C_0}
\end{equation}

And for the occupation numbers of BLG, it is simply
\begin{equation}
  n_{\mathrm{BLG}} = \rho_{\mathrm{BLG}} \cdot \mu
\end{equation}


Combining the above equations, we get
\begin{equation}
n_{\mathrm{tot}} = n_{\mathrm{BLG}} + n_\mathrm{I}  = \frac{e \rho_\mathrm{I} a (V_1 - V_2)
 + \rho_\mathrm{I} \mu}{1 + e^2 \rho_\mathrm{I} / C_0} + \rho_{\mathrm{BLG}} \cdot \mu 
\end{equation}

Combining Eq.(3) and Eq.(7), 
\begin{equation}
 \mu = \frac{2 C_{\mathrm{eff}} V_1 - e^2 a \tilde{\rho}_\mathrm{I} (V_1 - V_2)}{e \tilde{\rho}_\mathrm{I} + e \rho_{\mathrm{BLG}}} \nonumber \approx   e\left [  \left ( 2\frac{C_\mathrm{eff}}{C_0}-a \right ) V_1+a V_2 \right ] 
\end{equation}
\noindent   with $\tilde{\rho}_\mathrm{I} = \rho_\mathrm{I}/(1 + e^2 \rho_\mathrm{I}/ C_0) \approx {C_0}/{e^2} $.


Clearly, Eq.(8) depicts that, at a fixed $V_1$ in the SIC-phase, the chemical potential of the BLG is only a function of $V_2$, which explains why the parameter $D$ is related to an energy scale in the Landau quantizations in $D$-$B$ space, as in both BGL/CrOCl (Extended Data Figure 2) and MLG/CrOCl cases \cite{Submitted}. 

Indeed, a line profile along $V_1=1$ V is shown in Extended Data Figure 10d. According to the energy of BLG Landau levels, $E_\mathrm{\nu}=\hbar \omega_{c}\sqrt{|\nu|(|\nu|-1)}$, with $\nu$ the filling fractions, and $\omega_{c}=eB/m^{*}$ the cyclotron frequency. By adopting $m^{*} \sim 0.037 m_{e}$ \cite{mass_PRB_2009}, we can obtain $a=0.092$ in Eq.(8) by measuring the slope in Extended Data Figure 10e. Thus, the value of $C_\mathrm{eff}/C_{0}$ can be estimated to be around 0.055. However, precise value of $E_\mathrm{\nu}$ will require further investigations such as optical measurements, and we can only make estimations using the current conditions.

We now consider the built-in electrical field $E_\mathrm{Built-In}$ induced by the close-to-surface states $e n_\mathrm{I}/(d C_0)$, determined as
\begin{equation}
\frac{e^2 \widetilde{\rho_I} a (V_1 - V_2) + e
  \widetilde{\rho_I} \left[ \frac{2 C_{\mathrm{eff}} V_1 - e^2 a \tilde{\rho}_I
  (V_1 - V_2)}{e \tilde{\rho}_I + e \rho_{\mathrm{BLG}}} \right]}{d C_0} \approx
  \frac{2 e C_{\mathrm{eff}}}{d C_0} V_1
\end{equation}

With the $C_\mathrm{eff}/C_{0}$ ratio determined above, and the $d$ estimated to be 0.37 nm according to the TEM cross-section (Fig. 1c in the main text), we obtain an $E_\mathrm{Built-In}$ at the order of 10 V/nm at the largest $V_1$ available in our experiment. This is way higher than any state-of-the-art gate dielectric can reach, which further explains the wide gap opened in the BLG/CrOCl system. 


\section{\label{sec:level1}DATA AVAILABILITY}
The data that support the findings of this study will be available at the open-access repository Zenodo with a doi link, when accepted for publishing.

\section{\label{sec:level2}Code AVAILABILITY}
The codes used in theoretical simulations and calculations are available from the corresponding authors on reasonable request.

\section{\label{sec:level3}ACKNOWLEDGEMENT}
This work is supported by the National Key R$\&$D Program of China with Grants. 2019YFA0307800, 2017YFA0206301, 2018YFA0306900, 2019YFA0308402, and 2018YFA0305604. The authors acknowledge supports by the National Natural Science Foundation of China (NSFC) with Grants 11974357, U1932151, 11934001, 11774010, and 11921005. Growth of hexagonal boron nitride crystals was supported by the Elemental Strategy Initiative conducted by the MEXT, Japan ,Grant Number JPMXP0112101001, JSPS KAKENHI Grant Number JP20H00354 and A3 Foresight by JSPS. Jianhao Chen acknowledges support from Beijing Municipal Natural Science Foundation (Grant No. JQ20002).

\section{Author contributions}
Z.H. and Y.Y. conceived the experiment and supervised the overall project. K.Y., X.G., and Y.W. carried out device fabrications; X.G., K.Y., Y.W., T.Z., P.G., R.Z., S.C., J.C., Y.Y., and Z.H. carried out electrical transport measurements; P.G. and Y.Y. performed synthesis of bulk CrOCl crystals; Z.L., H.W., and X.L. carried out TEM characterizations; K.W. and T.T. provided high quality h-BN bulk crystals. Z.H., Y.Y., J.Z. and X.D. analysed the experimental data. X.D. carried out theoretical modellings. The manuscript was written by Z.H. with discussion and inputs from all authors.

\section{ADDITIONAL INFORMATION}
Competing interests: The authors declare no competing financial interests.

\end{document}